\documentclass[]{interact}

\usepackage{epstopdf} 
\usepackage{subfigure} 

\usepackage[numbers,sort&compress,merge]{natbib} 
\bibpunct[, ]{(}{)}{,}{n}{,}{,} 

\theoremstyle{plain}

\theoremstyle{definition}

\theoremstyle{remark}

\usepackage[version=3]{mhchem}
\usepackage{amsmath}
\begin{document}

\articletype{PREPRINT}

\title{Enhancing the formation of ionic defects to study the ice Ih/XI transition with molecular dynamics simulations}

\author{
\name{Pablo M.~Piaggi\textsuperscript{a}\thanks{CONTACT Pablo M.~Piaggi. Email: ppiaggi@princeton.edu} and Roberto Car\textsuperscript{b}}
\affil{\textsuperscript{a}Department of Chemistry, Princeton University, Princeton, NJ 08544, USA; \\ \textsuperscript{b}Department of Chemistry and Department of Physics, Princeton University, Princeton, NJ 08544, USA}
}

\maketitle

\begin{abstract}
Ice Ih, the common form of ice in the biosphere, contains proton disorder.
Its proton-ordered counterpart, ice XI, is thermodynamically stable below 72 K. 
However, even below this temperature the formation of ice XI is kinetically hindered and experimentally it is obtained by doping ice with KOH.
Doping creates ionic defects that promote the migration of protons and the associated change in proton configuration.
In this article, we mimic the effect of doping in molecular dynamics simulations using a bias potential that enhances the formation of ionic defects.
The recombination of the ions thus formed proceeds through fast migration of the hydroxide and results in the jump of protons along a hydrogen bond loop.
This provides a physical and expedite way to change the proton configuration, and to accelerate diffusion in proton configuration space.
A key ingredient of this approach is a machine learning potential trained with density functional theory data and capable of modeling molecular dissociation. 
We exemplify the usefulness of this idea by studying the order-disorder transition using an appropriate order parameter to distinguish the proton environments in ice Ih and XI.
We calculate the changes in free energy, enthalpy, and entropy associated with the transition.
Our estimated entropy agrees with experiment within the error bars of our calculation.
\end{abstract}


\section{Introduction}

By 1933 there was a growing body of evidence that calorimetric and spectroscopic determinations of the entropy of water did not agree with each other\cite{Giauque33,Giauque36}.
The crucial assumption of calorimetric measurements (later to be proven wrong) was that at 0 K the entropy of ice had to be zero.
It was also known from x-ray diffraction experiments that oxygen atoms in ice had the structure that we now call hexagonal diamond with each oxygen tetrahedrally coordinated\cite{Barnes29}.
However, Bernal and Fowler\cite{Bernal33} had pointed out in 1933 that it was conceivable that the orientation of water molecules in ice could be random while maintaining the periodic arrangement of oxygen atoms in the lattice.
They thus proposed the ice rules stating that, even in presence of disorder in the molecular orientations, each pair of oxygen nearest neighbors share only one hydrogen atom, and each oxygen atom has two strongly bonded and two weakly bonded hydrogen atoms.
The next step forward came from Pauling\cite{Pauling35} in 1935 who made a famous back-of-the-envelope calculation showing that the disorder of hydrogen atoms (protons) gave rise to a residual entropy in good agreement with the entropy inferred from spectroscopy.
Pauling's calculation reconciled spectroscopic and calorimetric measurements, and bolstered the hypothesis that ice had proton disorder.
Later, his estimate of the entropic contribution from proton disorder was subject to several refinements\cite{Nagle66,Dimarzio64}.
The solid form of water that we have so far simply referred to as ice, is nowadays called ice Ih since a large number of other ice polymorphs have since been discovered.

A proton ordered counterpart of ice Ih would be compatible with the third law of thermodynamics, yet for many years there was no experimental confirmation of its existence.
In the context of the study of the dielectric properties of ice, Bjerrum\cite{Bjerrum52}, Onsager\cite{Onsager60}, and others realized that strict adherence to the ice rules could kinetically hinder changes in proton configuration.
They then proposed mechanisms that could alter the proton configuration due to the presence of defects that locally violate the ice rules.
These defects are the Bjerrum and ionic defects (\ce{OH-} and \ce{H3O+}).
Onsager also surmised that doping ice with HF could increase the number of ionic and Bjerrum defects, and thus facilitate this process\cite{Onsager67}.
Tajima et al.~showed in 1982 that it was possible to obtain a proton-ordered form by doping ice with \ce{KOH} instead of \ce{HF}\cite{Tajima82}.
Subsequently, the structure of the proton ordered form was completely determined\cite{Leadbetter85,Howe89,Line96,Jackson97} and the polymorph is now called ice XI.
This polymorph is now believed to be ferroelectric\cite{Jackson95} and more stable than ice Ih below 72 K at ambient pressure\cite{Tajima82}.

Molecular simulations are a useful tool to obtain microscopic insight into physicochemical phenomena and several works have been devoted to the study of the ice Ih/XI transition.
Some studies aimed at calculating the residual entropy of ice Ih using simplified, yet structurally accurate, energy models\cite{Berg07,Herrero13}.
Sampling proton configurations in realistic molecular simulations is difficult as they suffer from the same limitation of experiments regarding the kinetic hindrance of changes in proton configuration.
In a pioneering simulation of ice in 1972, Rahman and Stillinger devised an algorithm to change the proton configuration by displacing protons along hydrogen-bond loops\cite{Rahman72}.
This type of algorithm has often been used in the context of Monte Carlo simulations\cite{Barkema98,Rick03,Rick05,Schonherr14}.
Recently, Lasave \textit{et al.} simulated the nucleation of the ferroelectric phase using replica-exchange Monte Carlo simulations of a spin-like Hamiltonian explicitly introducing ionic defects\cite{Lasave20}.

Another aspect of the ice Ih/XI transition that has been extensively studied computationally is the subtle variation in energy of different proton configurations using both semiempirical models\cite{Buch98,Hirsch04,Rick05} and density-functional theory (DFT)\cite{Hirsch04,Singer05,Tribello06,Raza11,Pamuk15}.
An issue that has received particular attention is the competition between two possible ground state proton configurations of hexagonal ice, namely the experimental ferroelectric $Cmc2_1$ configuration and the antiferroelectric $Pna2_1$ configuration proposed by Davidson and Morokuma\cite{Davidson84}.
Results suggest that semiempirical models are not able to describe properly the energetics of different proton configurations, while DFT is able to do so much better\cite{Hirsch04,Tribello06,Raza11}.
Nuclear quantum effects (NQE) have an influence on the energy variation between proton configurations and this has been studied by Pamuk \textit{et al.}\cite{Pamuk15}.

Here, we aim at studying the ice Ih/XI transition within molecular dynamics simulations.
We use classical simulations, thus, we do not address the isotope effect on the transition temperature ($\sim$4 K) found in experiments when substituting H with D\cite{Matsuo86}. 
Our work contains at least two elements of novelty.
First, we employ a machine learning potential trained with DFT data.
Therefore our model is able to describe the subtle variation in energy between different proton configurations and also the dissociation of the water molecule.
Second, we employ enhanced sampling methods in order to accelerate changes in proton configuration.
Our approach is based on the introduction of a bias potential that promotes the formation of oxygen atoms with three-fold proton coordination.
This procedure leads to proton jumps along hydrogen-bond loops that result in changes of the proton configuration.
At variance with other approaches, our method does not create permanent ionic defects in the structure of ice.
We then take advantage of the enhanced diffusion in proton configuration space to study the ice Ih/XI transition using an order parameter tailored for that purpose.
Finally, we calculate the polarization change during a defect induced transition from ice Ih to ice XI, and report the corresponding plot of the free energy versus polarization.

\section{Methods}

\subsection{Machine learning model for water}

One of the fundamental ingredients of our approach is a model of water able to describe the self-ionization and the proton transfer process.
We employ a machine learning model of water\cite{Gartner20} built using the DeePMD framework\cite{Zhang18,Zhang18end}.
The model is based on deep neural networks and was trained with DFT data obtained by adopting the strongly constrained and appropriately normed (SCAN) functional.
SCAN has been shown to provide a good description of the properties of liquid water and ice Ih\cite{Chen17}, including proton transfer processes in the liquid\cite{Chen18}.
The DeePMD model describes well water dissociation events as shown for instance in a recent study of the hydroxylation of a \ce{TiO2} surface exposed to water\cite{Andrade20}.
Proton transfer configurations in ice were not included in the training of the DeePMD model that we adopt here, which was taken from ref.~\citenum{Gartner20}.
Such training would likely be important to model proton transfer dynamics in ice, but it is not relevant in the current thermodynamic context, where these processes provide just a convenient path for transforming reversibly ice Ih into ice XI.

We used systems of 96 water molecules.
Simulations were performed in the isothermal-isobaric ensemble and we employed a timestep of 0.5 fs.
The temperature was kept constant using the stochastic velocity rescaling thermostat with a relaxation time of 0.1 ps.
The pressure was maintained at 1 bar using an anisotropic Parrinello-Rahman barostat.
Even though the sides of the box were allowed to change, the angles were fixed at 90$^o$.
All simulations were performed using the molecular dynamics engine LAMMPS\cite{Plimpton95} driven by the DeePMD-kit\cite{Wang18}.

\subsection{Enhanced sampling using a variational principle}

In this work we focus on two processes: the formation of ionic defects and the change in proton configuration to drive the transformation from ice Ih to ice XI.
These processes are rare events in the scale of a molecular dynamics simulation and we thus have to resort to techniques to accelerate them.
Enhanced sampling methods aim at increasing the probability to observe rare but crucial fluctuations.
Here we take advantage of the variationally enhanced sampling (VES) method introduced by Valsson and Parrinello\cite{Valsson14}.
This method is similar in spirit to umbrella sampling\cite{Torrie77} or metadynamics\cite{Laio02,Barducci08}, and is based on the introduction of a bias potential $V(\mathbf{s})$ that is a function of a set of collective variables (CVs) $\mathbf{s}$.
The CVs are continuous and differentiable functions of the atomic coordinates $\mathbf{R}$ and are chosen to describe the evolution of the process under study.
At the core of the VES method is a functional of $V(\mathbf{s})$, and it can be shown that the minimum of this functional is reached for
\begin{equation}
 V(\mathbf{s}) =  - F(\mathbf{s}) - \frac{1}{\beta} \log  p(\mathbf{s}),
\end{equation}
where $F(\mathbf{s}) = -(1/\beta) \log \int d\mathbf{R} \: \delta (\mathbf{s}-\mathbf{s}(\mathbf{R})) e^{-\beta [ U(\mathbf{R}) + \mathcal{P V}]}$ is the free energy, $\beta$ is the inverse temperature, $U(\mathbf{R})$ is the potential energy, $\mathcal{P}$ is the pressure, $\mathcal{V}$ is the volume, and $p(\mathbf{s})$ is the target probability distribution of the CVs.
$p(\mathbf{s})$ has to be chosen properly in order to surmount free energy barriers and increase the probability of observing rare configurations.
Typical choices of $p(\mathbf{s})$ are the uniform distribution $p(\mathbf{s})=\mathrm{const} $ and the well-tempered distribution\cite{Bonomi10,Valsson15} $p(\mathbf{s})= P(\mathbf{s})^{1/\gamma}$ where $\gamma \geq 1$ is the bias factor and $P(\mathbf{s})=\int d\mathbf{R} \: \delta (\mathbf{s}-\mathbf{s}(\mathbf{R})) e^{-\beta [U(\mathbf{R})+ \mathcal{P V}]}$ is the unbiased distribution of the CVs.

In practice, one has to expand the bias potential in basis functions and the expansion parameters are determined using an algorithm akin to stochastic gradient descent.
Further details can be found in ref.~\cite{Valsson14}.
We employed Legendre polynomials and the well-tempered target distribution.
Multiple walkers were used to improve the convergence of the bias potential.
The PLUMED enhanced sampling plugin\cite{Tribello14,Bonomi19,vescode} was used in tandem with LAMMPS in order to determine and apply the bias potential.

The bias potential alters the probability of observing a given configuration, and therefore the calculation of observables requires special attention.
A thorough discussion on the calculation of observables using the reweighting technique has been given in refs.~\cite{Valsson16review,Piaggi20,Invernizzi20}.

\subsection{Collective variable to enhance the formation of ionic defects}

The simplest ionic defects are the hydronium \ce{H3O+} and hydroxide \ce{OH-} ions.
More complex entities such as the Zundel ion \ce{H5O2+} and the Eigen ion \ce{H9O4+} are also possible, which, however, represent only limiting or ideal structures\cite{Marx99}.
In order to promote the formation of these defects it seems natural to consider the coordination number of each oxygen with hydrogen.

\begin{figure}[t]
\centering
\includegraphics[width=0.6\textwidth]{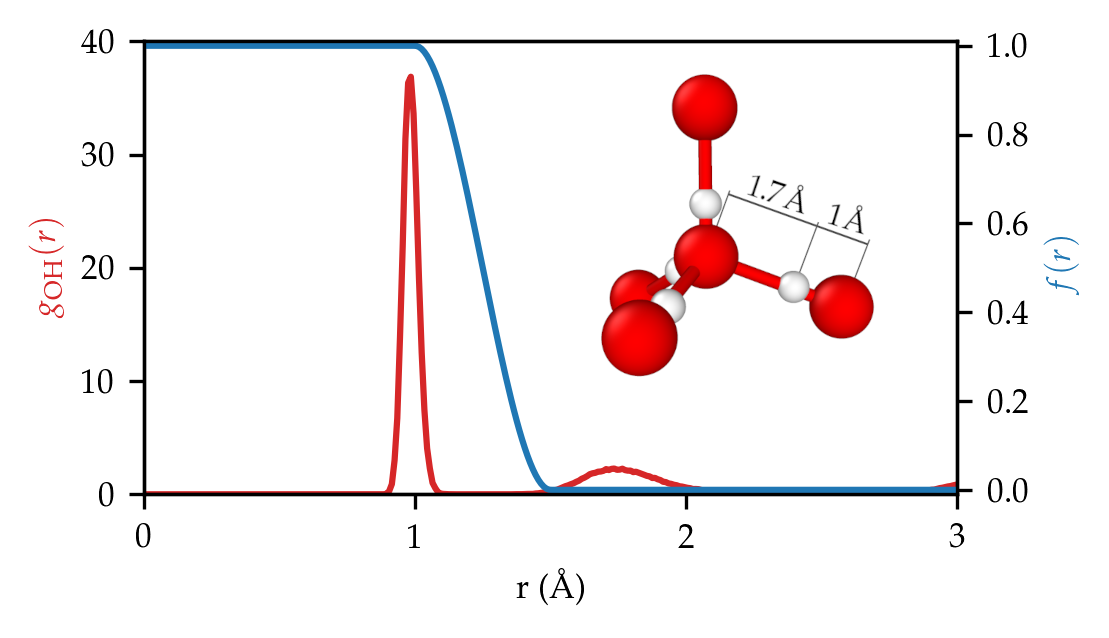}
\caption{\label{fig:plotgr} Radial distribution function $g_{\mathrm{\ce{OH}}}(r)$ between oxygen and hydrogen atoms and switching function $f(r)$ used to calculate the coordination number of hydrogen with oxygen. The environment around an oxygen atom shows a case in which the ice rules are satisfied, i.e. there are two hydrogen atoms at $\sim$ 1 \AA\ and another two hydrogen atoms at $\sim$ 1.7 \AA. Therefore the coordination number computed using $f(r)$ is $\sim$ 2 in this case. Oxygen and hydrogen atoms shown in red and white, respectively.}
\end{figure}

The coordination number of the $i-$th oxygen is,
\begin{equation}
    n_i = \sum\limits_{j \in \mathrm{H}} f(r_{ij}),
\end{equation}
where $r_{ij}$ is the norm of the distance between oxygen atom $i$ and hydrogen atom $j$, and $f(r_{ij})$ is a continuous and differentiable switching function that is one at short distance and zero at distances larger than some radius $r_{max}$.
For our simulations we have chosen,
\begin{equation}
f(y)=
\begin{cases}
    1 \quad \mathrm{if} \quad y<0 \\
    (y-1)^2(1+2y) \quad \mathrm{if} \quad 0<y<1 \\
    0 \quad \mathrm{if} \quad y>1\\
\end{cases}
,
\end{equation}
where $y=(r-r_0)/(r_{max}-r_0)$, $r_0$ is a distance below which all hydrogen atoms are considered neighbors, and beyond the distance $r_{max}$ all hydrogen atoms are ignored.
For the model studied here, the first and second peaks of the O-H radial distribution function $g_{\mathrm{\ce{OH}}}(r)$ are located at around $1$ and $1.7$ \AA\ as shown in Figure \ref{fig:plotgr}.
These peaks correspond to the two strongly bonded and two weakly bonded protons considered in the ice rules (see oxygen environment in Figure \ref{fig:plotgr}).
Based on this information we have used $r_0=1$ \AA\  and $r_{max}=1.5$ \AA.
The switching function $f(r)$ based on this choice is shown in Figure \ref{fig:plotgr}.
Therefore, $n_i\approx 2$ if the ice rules are satisfied.

By promoting changes in the coordination number we will create local deviations from the ice rules.
We now consider the number of oxygen atoms that have a coordination number larger than two.
This is equivalent to the number of oxygen atoms with three-fold hydrogen coordination since coordination numbers larger than three were not observed.
Thus, we use the notation $N_{\mathrm{\ce{H3O}}}$.
This quantity can be written as,
\begin{equation}
    N_{\mathrm{\ce{H3O}}} =  \sum_{i \in \mathrm{O}} \theta (n_i-n_0)
    \label{eq:def_NH3O_not_cont}
\end{equation}
where $\theta(n_i-n_0)$ is a characteristic function such that $\theta(n_i-n_0)=0$ if $n_i<n_0$ and $\theta(n_i-n_0)=1$ if $n_i>n_0$.
We choose $n_0=2.5$ in order to count the number of oxygen atoms that have a coordination number larger than two.
$N_{\mathrm{\ce{H3O}}}$ can also be made continuous and differentiable by replacing $\theta(n_i-n_0)$ by a suitable switching function,
\begin{equation}
    N_{\mathrm{\ce{H3O}}} =  \sum_{i \in \mathrm{O}} \frac{x^p - x^q }{1-x^q},
    \label{eq:def_NH3O}
\end{equation}
where $N$ is the total number of oxygen atoms, $x=n_i/n_0$, and the exponents $p$ and $q$ control the steepness of the switching function. The choice $p=24$,  $q=48$ provides a good compromise between a clear separation of oxygen atoms with $n_i=2$ and $n_i=3$, and a smooth transition from zero to one in the switching function.
We shall use this definition of $N_{\mathrm{\ce{H3O}}}$ as a collective variable in the VES formalism to promote the formation of ionic defects with excess protons.
The formation and subsequent migration of the ionic defects will lead to changes in proton configuration.
We note that more complex collective variables to study acid-base equilibria have been introduced by Grifoni \textit{et al.}\cite{Grifoni19}.
However, we found that the simple collective variable described here was sufficient to promote changes in proton configuration in hexagonal ice.

\subsection{Collective variable to drive the ice Ih/XI transition}

Now that we have discussed how to promote changes in proton configuration, we turn our attention to  the transformation of ice Ih into ice XI.
In this section we describe an order parameter that distinguishes ice Ih from ice XI.
Both polymorphs share the same structure of the oxygen atoms yet differ in the position of the protons.
The positions of the protons in ice XI are uniquely defined since it is an ordered structure.
However, the hexagonal diamond structure of the oxygen sublattice has four different environments (the number of molecules in the primitive cell of ice Ih is four) and we will have to characterize each of them.
Furthermore, ferroeletric ice XI has two possible orientations with opposite dipole moments within the same sublattice of oxygen atoms.
This results from the mirror symmetry with respect to the basal plane.
The symmetry between the two possible orientations leads to spontaneous symmetry breaking, i.e. the two states are degenerate and the system chooses one of them during the transformation.  

\begin{figure}[t]
\centering
\includegraphics[width=\textwidth]{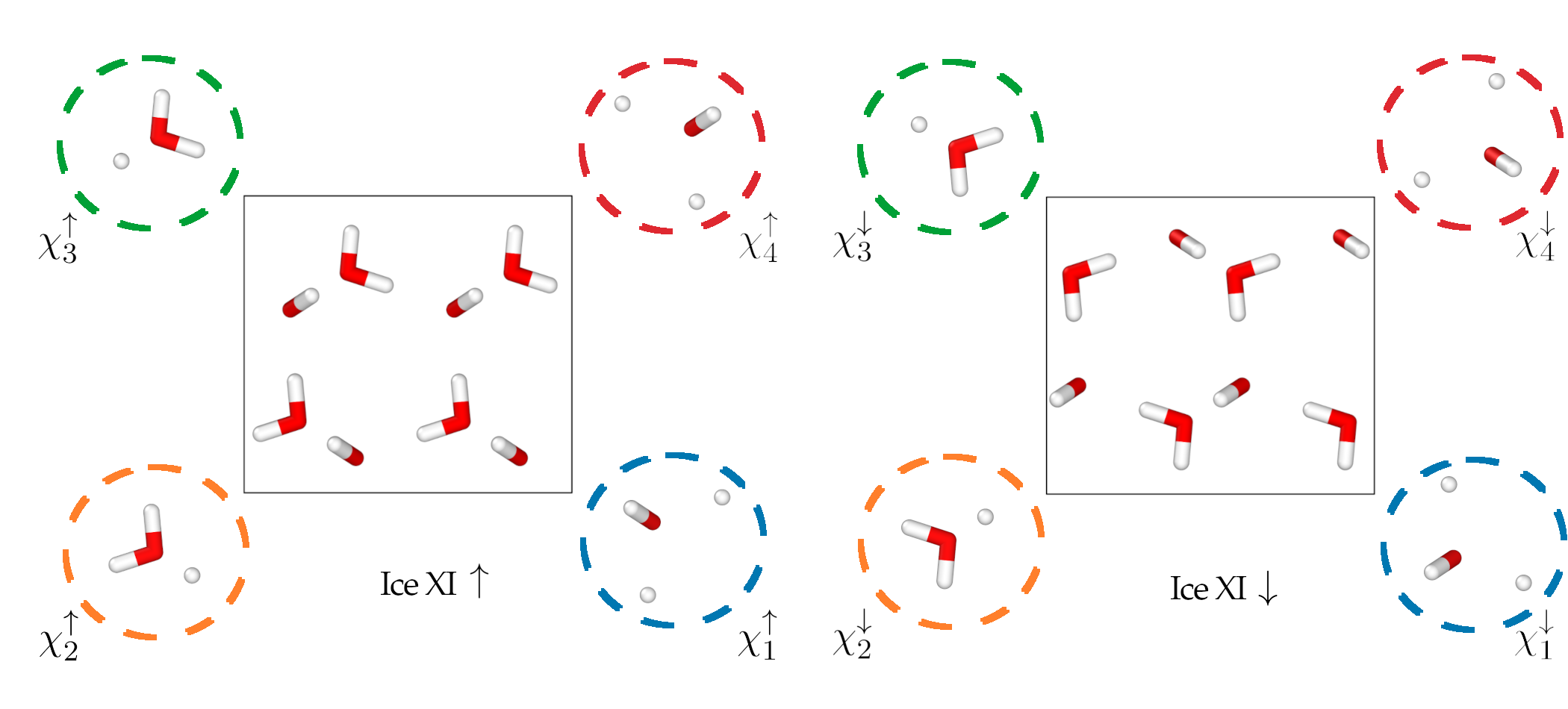}
\caption{\label{fig:envs} Environments around oxygen atoms in ice XI. There are four environments $X^{\uparrow} = \{ \chi_1^{\uparrow},\chi_2^{\uparrow},\chi_3^{\uparrow},\chi_4^{\uparrow} \} $ for the up orientation ($\uparrow$) of the dipole moment, and four environments $X^{\downarrow} = \{ \chi_1^{\downarrow},\chi_2^{\downarrow},\chi_3^{\downarrow},\chi_4^{\downarrow} \} $  for the down orientation ($\downarrow$) of the dipole moment.}
\end{figure}

\begin{figure}[t]
\centering
\includegraphics[width=0.6\textwidth]{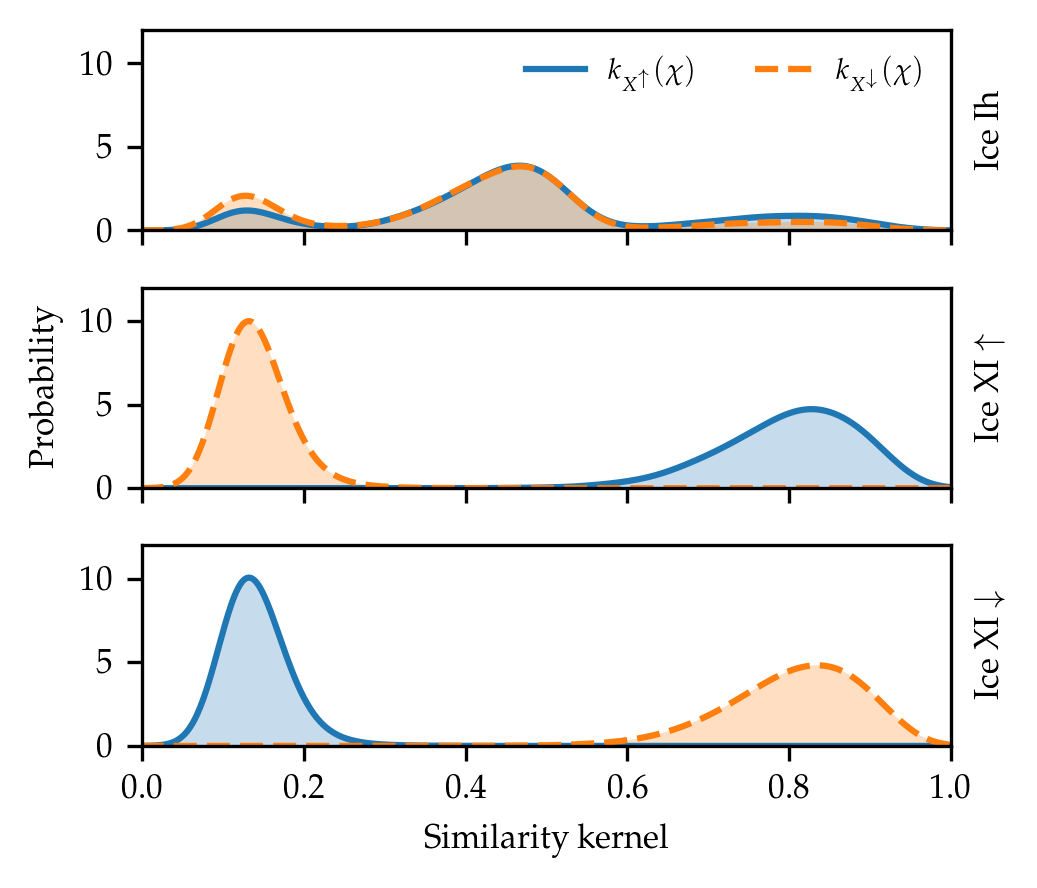}
\caption{\label{fig:plotDistributions} Distribution of the similarity kernels $k_{X^{\uparrow}}(\chi)$ and $k_{X^{\downarrow}}(\chi)$ in ice Ih, ice XI $\uparrow$, and ice XI $\downarrow$ at 300 K. }
\end{figure}

In order to describe the local structure of ice XI we shall use four environments $X^{\uparrow} = \{ \chi_1^{\uparrow},\chi_2^{\uparrow},\chi_3^{\uparrow},\chi_4^{\uparrow} \} $ for the up orientation ($\uparrow$) of the dipole moment, and four environments $X^{\downarrow} = \{ \chi_1^{\downarrow},\chi_2^{\downarrow},\chi_3^{\downarrow},\chi_4^{\downarrow} \} $  for the down orientation ($\downarrow$).
These environments are shown in Figure \ref{fig:envs}.
Using these environments we can construct a similarity measure\cite{Bartok13} using a similarity kernel\cite{Piaggi19b}.
This type of similarity measure has been used to study the crystallization of simple metals\cite{Piaggi19b}, of ice Ih\cite{Piaggi20}, and of several polymorphs of gallium\cite{Niu20}.
We restrict the following discussion to the $\uparrow$ variant of ice XI although the same discussion applies to the $\downarrow$ variant.
We define the kernel between the four reference environments $\chi_l^{\uparrow} \in X^{\uparrow}$ and a generic environment $\chi$,
\begin{equation}
    k_{\chi_l^{\uparrow}}(\chi) = \int \rho_{\chi_l^{\uparrow}}(\mathbf{r}) \rho_{\chi}(\mathbf{r}) \: d\mathbf{r},
    \label{eq:kernel1}
\end{equation}
where $\rho_{\chi_l^{\uparrow}}(\mathbf{r})$ and $\rho_{\chi}(\mathbf{r})$ are the atomic densities corresponding to the environments $\chi_l^{\uparrow}$ and $\chi$, respectively.
We shall represent the density using sums of Gaussians with spread $\sigma$, centered at the neighbors' positions, and rewrite the kernel as:
\begin{equation}
    k_{\chi_l^{\uparrow}}(\chi) = \frac{1}{n} \sum_{i \in \chi_l^{\uparrow}} \sum_{j \in \chi} \exp\left(-\frac{|\mathbf{r}_i^l-\mathbf{r}_j|^2}{4 \sigma^2}\right),
    \label{eq:kernel2}
\end{equation}
where $n$ is the number of neighbors in the environment $\chi_l^{\uparrow}$, and $\mathbf{r}_i^l$ and $\mathbf{r}_j$ are the positions of the neighbors in environments $\chi_l^{\uparrow}$ and $\chi$, respectively.
In Eq.\ \eqref{eq:kernel2} we have included a normalization such that $k_{\chi_l^{\uparrow}}(\chi_l)=1$.
Now we have four similarity kernels that allow us to identify whether a given environment is compatible with one of the four proton environments in ice XI.
A similarity measure between a given environment and any of the four proton environments in the $\uparrow$ variant of ice XI is,
\begin{equation}
    k_{X^{\uparrow}}(\chi) = \max \{k_{\chi_1^{\uparrow}}(\chi),k_{\chi_2^{\uparrow}}(\chi),k_{\chi_3^{\uparrow}}(\chi),k_{\chi_4^{\uparrow}}(\chi) \}.
    \label{eq:kernel_multi1}
\end{equation}
In Figure \ref{fig:plotDistributions} we show the distribution of $k_{X^{\uparrow}}(\chi)$ and $k_{X^{\downarrow}}(\chi)$ in ice Ih, ice XI $\uparrow$, and ice XI $\downarrow$ at 300 K.
Since ice Ih can have any environment compatible with the ice rules, there is a variety of environments that give rise to three peaks in the distribution of $k_{X^{\uparrow}}(\chi)$ and $k_{X^{\downarrow}}(\chi)$.
This is expected since Ih might contain some of the environments that exist in ice XI.
The interpretation of the distribution in ice XI is straightforward.
In ice XI $\uparrow$ the kernel $k_{X^{\uparrow}}(\chi)$ recognizes all environments as compatible with those of ice XI $\uparrow$ and the kernel $k_{X^{\downarrow}}(\chi)$ founds little or no similarity with the environments of ice XI $\uparrow$.
The converse is true for the distributions in ice XI $\downarrow$.

These kernels characterize the environment around a single atom.
However, for a system with $N$ water molecules there will be $N$ oxygen-hydrogen environments $\chi^1,\chi^2,...,\chi^N$.
We consider the average value of the kernel,
\begin{equation}
    \bar{k}^{\uparrow} = \frac{ \sum\limits_{i=1}^N k_{X^{\uparrow}}(\chi^i)}{N},
    \label{eq:op_avg}
\end{equation}
and the number of environments consistent with the $\uparrow$ variant of ice XI,
\begin{equation}
    n_{\mathrm{XI}}^{\uparrow} = \{\mathrm{number \: of} \: \chi^i \: : \: k_{X^{\uparrow}}(\chi^i)>\kappa \},
    \label{eq:op_num}
\end{equation}
where $\kappa$ is a threshold separating the values of $k_{X^{\uparrow}}(\chi^i)$ consistent with ice Ih and those consistent with ice XI.

These global order parameters are restricted to one of the variants of ice XI.
We can also define an order parameter akin to the one used in Landau's theory of phase transitions\cite{LandauBookStatisticalPhysics},
\begin{equation}
    \bar{k} = \bar{k}^{\uparrow} -\bar{k}^{\downarrow}
    \label{eq:op_bar_k}
\end{equation}
that satisfies $\bar{k} \approx 0$ in ice Ih (the disordered phase), $\bar{k} > 0$ in the $\uparrow$ variant of ice XI, and $\bar{k} < 0$ in the $\downarrow$ variant of ice XI.
This order parameter is able to distinguish the two variants of ice XI.
However, both variants are equivalent by symmetry and therefore thermodynamic properties, such as, e.g. the free energy, as a function of $\bar{k}$ must be even functions.
It is convenient to include this symmetry in the order parameter using
\begin{equation}
    | \bar{k} | = |\bar{k}^{\uparrow} -\bar{k}^{\downarrow}|
    \label{eq:op_abs_bar_k}
\end{equation}
that satisfies $|\bar{k}| \approx 0$ in ice Ih, and $|\bar{k}| > 0$ in ice XI.
We shall see that better resolution of the local changes in structure can be achieved using
\begin{equation}
    |n_{\mathrm{XI}}| = |n_{\mathrm{XI}}^{\uparrow} - n_{\mathrm{XI}}^{\downarrow}|.
\end{equation}
It is easy to see that $|n_{\mathrm{XI}}| \approx 0$ in ice Ih, and $|n_{\mathrm{XI}}| \approx N$ in ice XI.
We will use a continuous and differentiable version of $n_{\mathrm{XI}}$ to construct a bias potential with the VES framework (see reference \cite{Piaggi20} for details).

\subsection{Polarization}

We calculated the change in the macroscopic polarization of the system according to the modern theory of polarization\cite{Resta92,King93}, adopting the formulation in terms of Wannier centers\cite{Resta94}. Then, in our system the dipole moment $\boldsymbol\mu$ of the simulation cell is given, modulo a quantum, by:
\begin{align}
\boldsymbol\mu = \sum_{O}  6 \: e \: \mathbf{r}_O + \sum_{H} e \: \mathbf{r}_H + \sum_{C} (-8\:e) \: \mathbf{w}_C
\label{eq:dipole}
\end{align}
where $e$ is the elementary charge, $\mathbf{r}_O$, $\mathbf{r}_H$, and $\mathbf{w}_C$ are the positions of the oxygen atoms, hydrogen atoms, and Wannier centroids, and the indices $O,H$ and $C$ run over the hydrogen atoms, oxygen atoms, and Wannier centroids, respectively.
The charge of each oxygen atom is $+6e$ since we adopt a pseudopotential fomulation in which the atomic nuclei are replaced by ions having the charge of the nucleus minus that of the core electrons.
The Wannier centroid used in Eq.~\eqref{eq:dipole} is defined as the geometric center of the four maximally localized Wannier functions\cite{Marzari97} closest to each oxygen atom.
The charge of the Wannier centroids is $-8e$ in Eq.~\eqref{eq:dipole} due to the fact that within our pseudopotential formulation there are four orbitals with two electrons each, that can be uniquely associated to each oxygen atom.
The dependence of the Wannier centroids on the coordinates of the atoms in the system was described by a deep neural network (DNN) as described in ref.~\citenum{Zhang20b}.
The DNN that we use here was trained with SCAN data for liquid water and ice, as reported in ref.~\citenum{Sommers20}.  
The DeePMD-kit\cite{Wang18} was used to train the DNN and to calculate the position of the Wannier centroids for a given atomic configuration.
The polarization $P$ along the $c$ crystallographic axis is defined as the dipole moment along that axis per unit volume.

\section{Results and discussion}

\begin{figure}[t]
\centering
\includegraphics[width=0.6\textwidth]{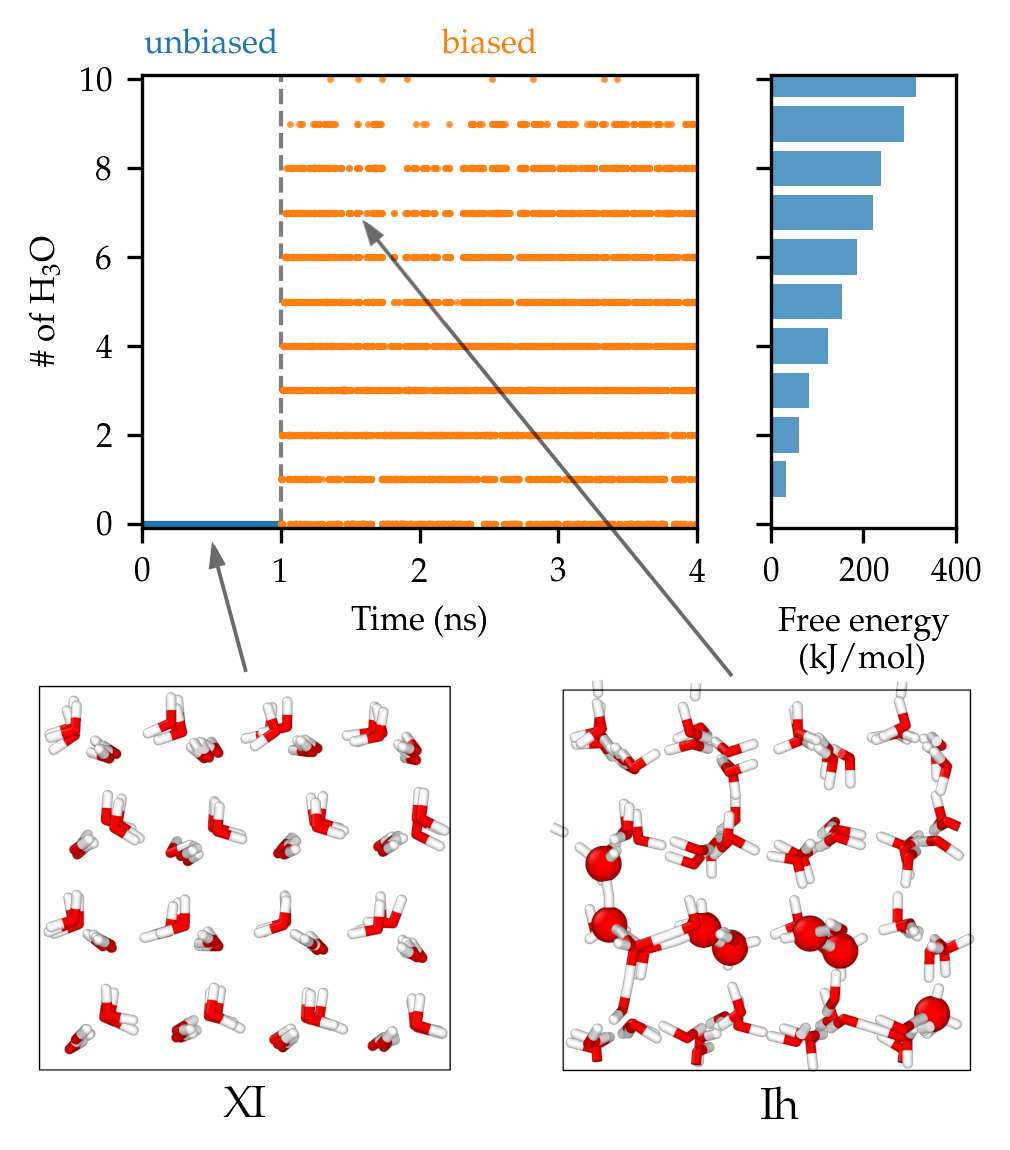}
\caption{\label{fig:plotNH3O} Number of \ce{H3O}-like defects as a function of simulation time. After 1 ns of unbiased simulation, a bias potential is introduced to enhance the formation of ionic defects. Trajectories of four multiple walkers are shown consecutively for the biased portion of the results. The free energy as a function of the number of \ce{H3O}-like defects is also shown. The initial ice XI configuration is shown below and the proton ordered structure can be seen. After 750 ps of biased simulation the configuration shows proton disorder compatible with ice Ih. Oxygen atoms are shown in red and hydrogen atoms are shown in white. \ce{H3O} defects are shown with red spheres.}
\end{figure}

We performed a 1 ns long unbiased molecular dynamics simulation of ice XI at 300 K.
In Figure \ref{fig:plotNH3O} we show the number of \ce{H3O}-like defects in the system as a function of simulation time.
The number of \ce{H3O}-like defects can be calculated using Eq.~\eqref{eq:def_NH3O} and replacing $f(r)$ with a step function at 1.25 \AA.
We observed that the number of \ce{H3O}-like defects remained close to zero throughout the simulation, implying that the ice rules were always satisfied and that no ionic defects were spontaneously created.

We then introduced a bias potential as a function of the collective variable $N_{\mathrm{\ce{H3O}}}$ defined in  Eq.~\eqref{eq:def_NH3O}.
The bias potential was optimized to target the well-tempered distribution with bias factor 50.
After 250 ps the bias was quasi-static, well-converged and the system explored a number of \ce{H3O}-like defects varying from zero to around ten as shown in Figure \ref{fig:plotNH3O}.
An inspection of the proton configuration at around 750 ps after the bias potential was introduced, shows that the ordered proton configuration of ice XI has changed to a disordered configuration (see snapshot in Figure \ref{fig:plotNH3O}).
We also observed that the defects created by the bias potential are not isolated \ce{H3O}-like entities but rather they involve larger numbers of oxygen atoms along hydrogen bond loops or strings (see snapshot in Figure \ref{fig:plotNH3O}).
From the simulation we can also compute the free energy associated with the creation of different numbers of \ce{H3O}-like defects.
The formation of a single \ce{H3O} defect has a free energy cost of around 30 kJ/mol.
This free energy should not be confused with the formation free energy of a \ce{H3O+} - \ce{OH-} pair, since in this case the two ions must be well separated. 

\begin{figure}[t]
\centering
\includegraphics[width=\textwidth]{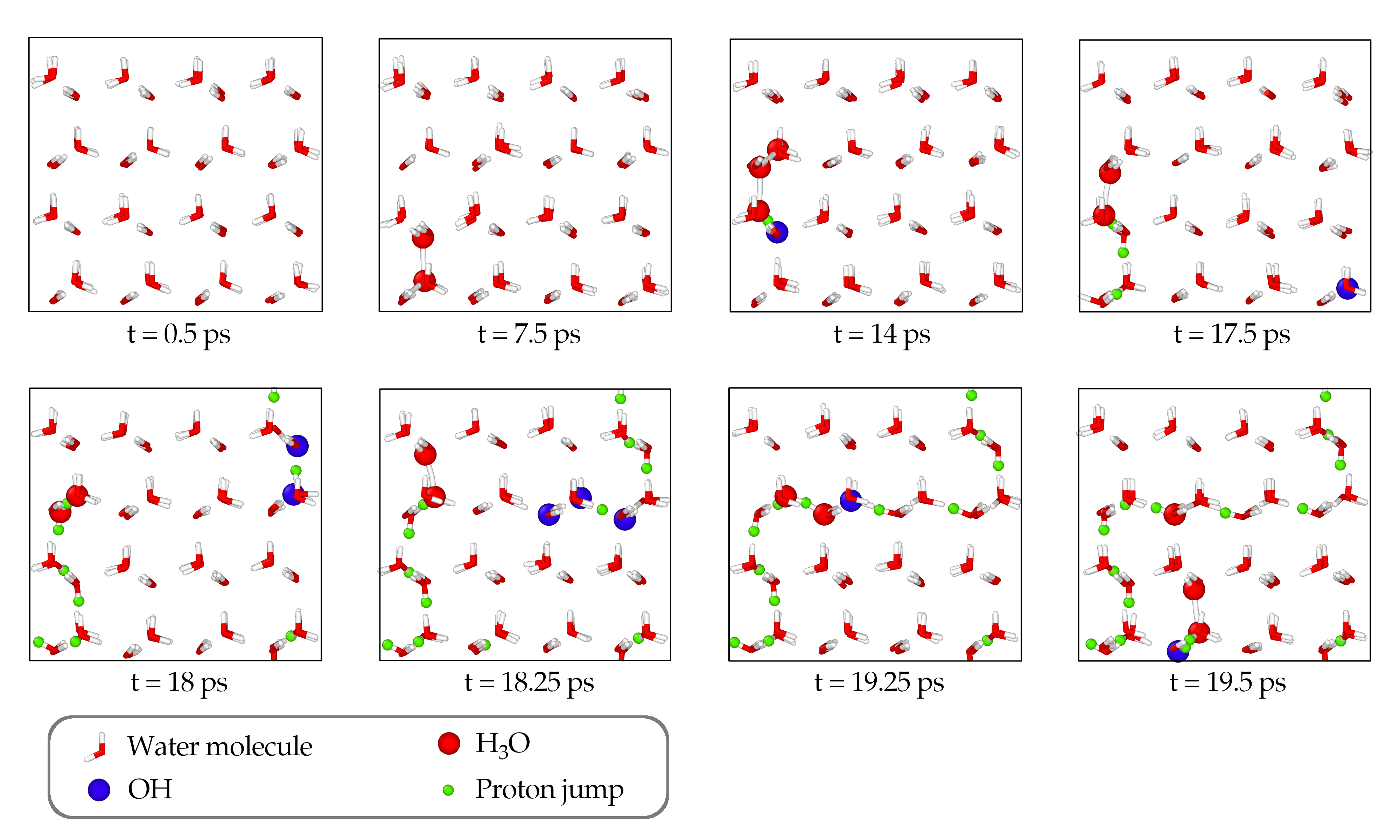}
\caption{\label{fig:sequence} Sequence of configurations that illustrate the mechanism by which changes in proton configurations are achieved. Oxygen and hydrogen atoms are shown in red and white, respectively. \ce{H3O} and \ce{OH} defects are shown with red and blue spheres. Protons that have jumped with respect to their initial configurations are shown in green.}
\end{figure}

We now turn to study the mechanism by which changes in proton configuration are achieved.
To this end we analyzed the first proton jumps along a hydrogen-bond loop observed during the biased simulation.
This event is shown with snapshots of the system's configuration in Figure \ref{fig:sequence}.
Initially, the proton configuration is ordered and corresponds to ice XI.
After 7.5 ps there are fluctuations that promote the sharing of a proton by two oxygen atoms, yet there is not a clear spatial separation of charge in the system.
The change in proton configuration is triggered at 14 ps by the formation of an \ce{H7O3+} cation and an \ce{OH-} anion that leads to the first proton jump and charge separation.
Further proton jumps occur in the next few ps as the \ce{OH-} propagates along a hydrogen bond chain.
After the initial proton jump, the \ce{H7O3+} seems to transform into a \ce{H5O2+} (Zundel) cation.
However, the initial formation of the \ce{H7O3+} cation does not seem to be crucial since in other events we observed only the Zundel ion.
The snapshot at 18.25 ps shows evidence of collective behavior in the diffusion of the \ce{OH-} anion since at least three water molecules seem to be involved in a concerted proton jump.
At 19.5 ps no ion is observed along this hydrogen-bond loop, but a new pair of Zundel and \ce{OH-} has formed elsewhere in the simulation box.
Further events follow a similar mechanism.
While most proton jumps are a consequence of the diffusion of the \ce{OH-} anion, the diffusion of the Zundel cation results in two proton jumps.
A total of 16 proton jumps occur in a time span of around 5.5 ps.
We also analyzed the displacement of the protons and found that after 1 ns around 60 \% of the protons have been displaced from their original positions.

The analysis above implies a higher mobility of the \ce{OH-} ion than of the \ce{H3O+} ion in ice Ih.
As a matter of fact, we can calculate the ratio of the diffusion coefficients from the limited number of jumps described above to be $D_{\textrm{\ce{OH-}}}/D_{\textrm{\ce{H3O+}}} \approx 6$.
This result would be in stark contrast with the situation in liquid water.
There is significant evidence that the diffusion of \ce{H3O+} in liquid water is almost two times faster than the diffusion of \ce{OH-}\cite{Halle83,Sluyters10}, i.e. $D_{\textrm{\ce{OH-}}}/D_{\textrm{\ce{H3O+}}} \approx 0.5$.
This behavior in liquid water has been attributed to a stabilized hypercoordinated solvation structure around \ce{OH-}\cite{Lee11,Chen18}.
In ice Ih the formation of such a structure is not allowed as a consequence of the crystalline structure of oxygen atoms.
As far as we know, to date there is no experimental evidence that $D_{\textrm{\ce{OH-}}}>D_{\textrm{\ce{H3O+}}}$ in ice.
The availability of data on the diffusion coefficient of \ce{OH-} and \ce{H3O+} in ice is limited compared to corresponding data for liquid water\cite{PetrenkoIce}.
The limited data is a result of experimental difficulties, and the fact that the charge transport in ice supposedly involves not only the ionic defects \ce{OH-} and \ce{H3O+}, but also the Bjerrum L and D defects\cite{PetrenkoIce}.
As a result of these issues, the estimations of the steady-state conductivity and of the diffusion coefficients of ionic defects can vary over several orders of magnitude\cite{PetrenkoIce}.
From the event that we have analyzed, we can also calculate approximately the jump frequencies and diffusion coefficients of \ce{OH-} and \ce{H3O+} that we report in Table \ref{tbl:jf_dc}.
Onsager and Dupuis\cite{Onsager60} estimated the sum of the jump frequencies, but their result relies on disputed measurements of the steady-state conductivity\cite{PetrenkoIce}.
\begin{table}[b]
\small
  \caption{Jump frequency $\nu$ and diffusion coefficient $D$ of ionic defects at 300 K (based on very limited information of one \ce{OH-}-\ce{H3O+} recombination event). The diffusion coefficient is calculated using the formula $D=\nu d^2 / 6$ with $d=2.76$ \AA\ the interoxygen distance.}
  \label{tbl:jf_dc}
  \begin{tabular*}{\textwidth}{@{\extracolsep{\fill}}lll}
    \hline
     & $\nu$ (ps$^{-1}$) & $D$ ($10^{-8}$ m$^2$/s) \\
    \hline
     \ce{OH-} & $\sim 2.5$ & $\sim 3.2$ \\
     \ce{H3O+} & $\sim 0.4$ & $\sim 0.5 $\\
     \ce{OH-} + \ce{H3O+} & $\sim 2.9$ & $\sim 3.7$\\
    \hline
  \end{tabular*}
\end{table}

The phase transformation from ice XI to ice Ih entails a change in the total dipole moment.
However, flipping all protons along a hydrogen bond loop contained in the simulation box, conserves the dipole moment of the system.
In our simulations we observe also winding loops, i.e. loops that end in a different periodic image of the system.
Flipping protons along winding loops does change the total dipole moment and allows the system to undergo the ice XI/Ih transition.
Even though real systems do not have periodic boundary conditions, one can surmise a similar mechanism for changes in proton configuration in real systems provided that there are internal surfaces or other defects that can act as sources and sinks of ionic defects.

\begin{figure}[t]
\centering
\includegraphics[width=0.6\textwidth]{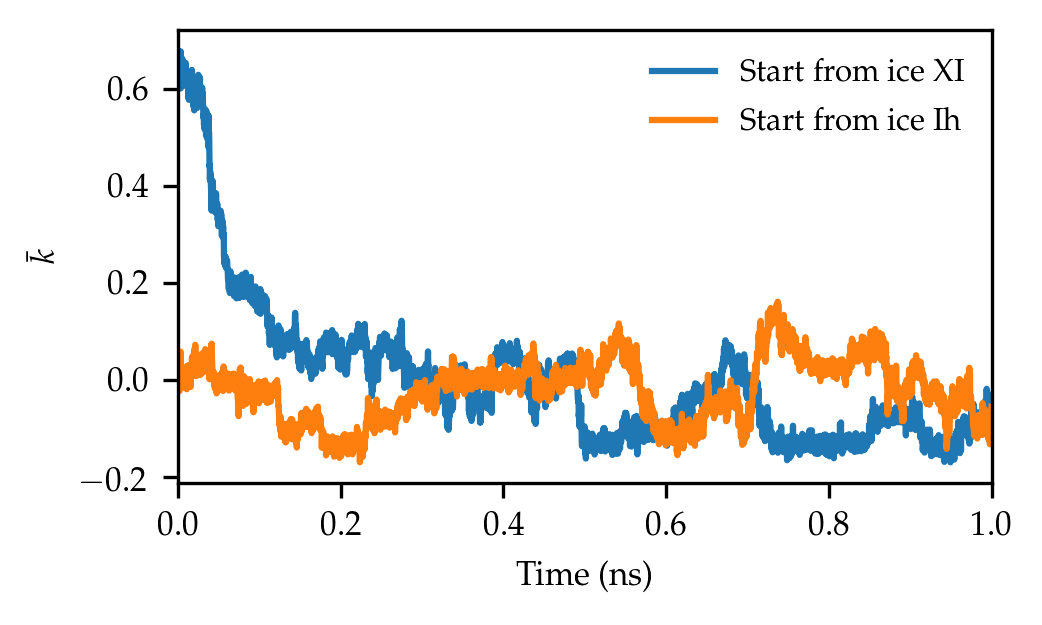}
\caption{\label{fig:plotbark} Order parameter $\bar{k}$ as a function of simulation time at 300 K. The biased simulation that starts from the $\uparrow$ variant of ice XI evolves to a disorder proton structure compatible with ice Ih after around 200 ps. $\bar{k}$ in the biased simulation starting from ice Ih does not see appreciable changes in its value. }
\end{figure}

We can also use the order parameter $\bar{k}$ defined in Eq.~\eqref{eq:op_bar_k} to analyze the change from the ordered proton configuration of ice XI to a disordered proton configuration.
In Figure \ref{fig:plotbark} we show $\bar{k}$ as a function of simulation time in a biased simulation starting from ice XI and another one starting from ice Ih.
The biased simulation that starts from ice Ih has a $\bar{k} \sim 0$ throughout the simulation in spite of the changes in proton configuration.
On the other hand, the biased simulation that starts from the $\uparrow$ variant of ice XI has an initial value of around $\bar{k} \sim 0.65$ but within 200 ps reaches $\bar{k} \sim 0$ and preserves that value throughout the rest of the simulation.
An unbiased simulation would show no changes in $\bar{k}$ since no changes in proton configurations occur in this case.

\begin{figure}[t]
\centering
\includegraphics[width=0.6\textwidth]{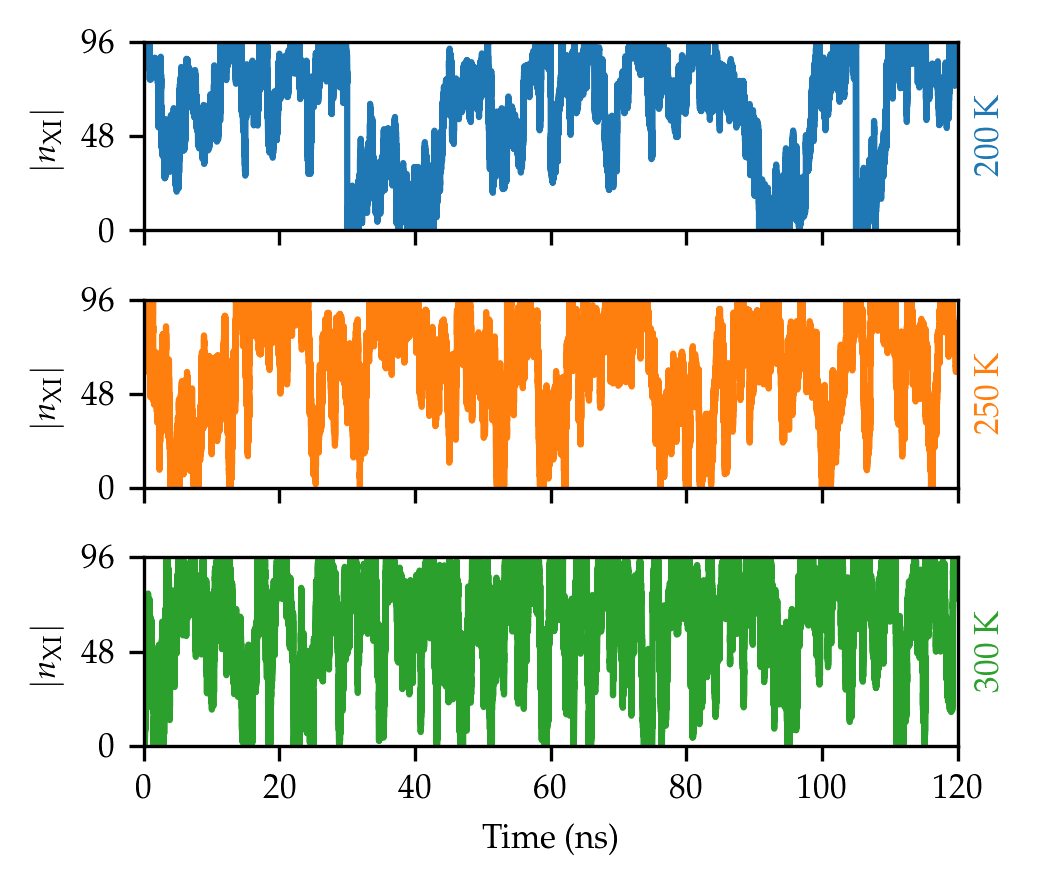}
\caption{\label{fig:plotnXI} Order parameter $|n_{\mathrm{XI}}|$ as a function of simulation time. Simulations at 200, 250, and 300 K are shown. The trajectories of eight multiple walkers are shown consecutively. }
\end{figure}

The analysis in Figure \ref{fig:plotbark} shows also that promoting changes in the proton configuration by itself does not suffice to observe reversible transitions between ice Ih and ice XI.
For this reason, we introduced another bias potential using VES and $|n_{\mathrm{XI}}|$ as CV.
In Figure \ref{fig:plotnXI} we show $|n_{\mathrm{XI}}|$ as a function of time at 200, 250, and 300 K.
The bias potential allows the system to explore reversibly ordered and disordered proton configurations.
$|n_{\mathrm{XI}}| \sim 0$ corresponds to ice Ih and $|n_{\mathrm{XI}}| \sim 96$ corresponds to ice XI.
The number of transitions between the ordered and disordered proton configurations falls sharply as the temperature is lowered from 300 K to 200 K.
We attribute the drop in the number of transitions  to the slower creation of ionic defects at low temperature.
For this reason we did not study lower temperatures.
It is possible that changing the definition of the switching functions used to define the collective variables could help to further enhance the creation of ionic defects.
An approach like parallel tempering\cite{Sugita99} could also be used to study lower temperatures.

\begin{figure}[t]
\centering
\includegraphics[width=0.6\textwidth]{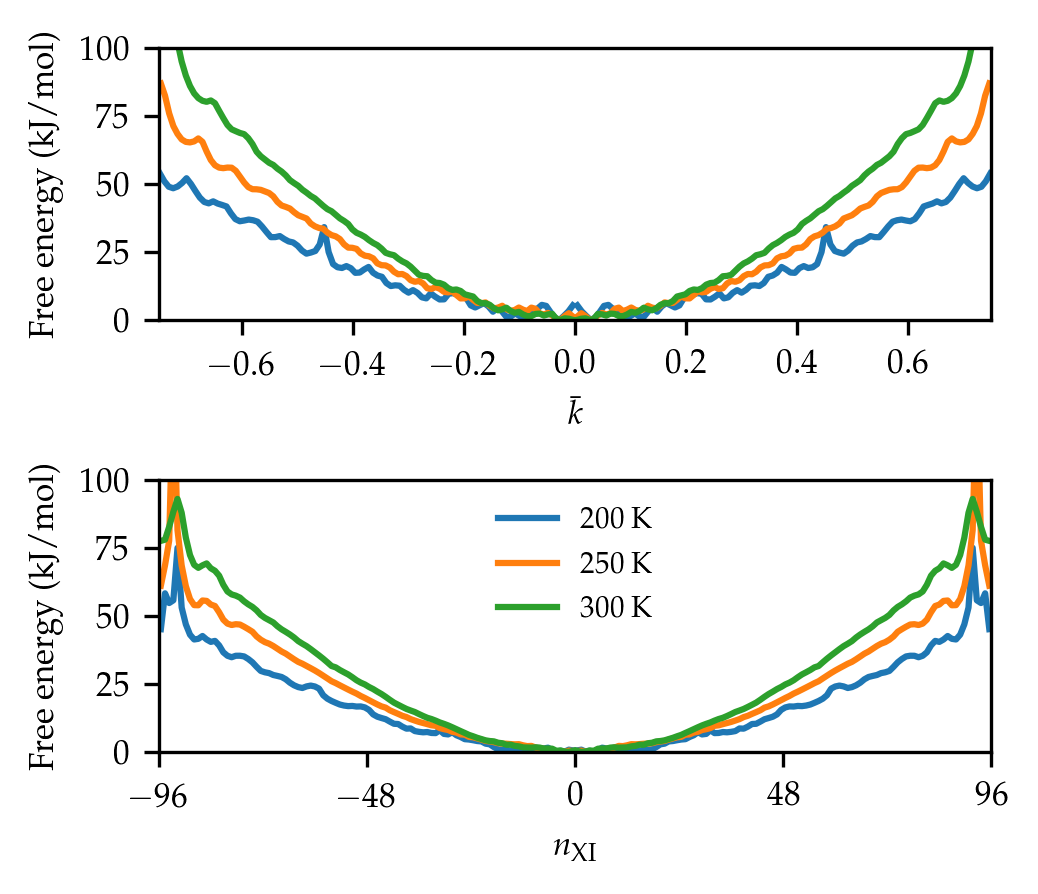}
\caption{\label{fig:plotLandau} Free energy as a function of the order parameters $\bar{k}$ and $n_{\mathrm{XI}}$.}
\end{figure}

From the simulations described above we can calculate the free energy as a function of $\bar{k}$ and $n_{\mathrm{XI}}$ using reweighting.
These functions are plotted in Figure \ref{fig:plotLandau} and bear similarity to Landau free energy profiles above the critical temperature.
We have enforced the even parity of the free energy required by symmetry.
The minimum of the free energy is located at $n_{\mathrm{XI}} \sim 0$ and $\bar{k} \sim 0$.
Thus, ice Ih is the most stable phase at the temperatures studied here.
In the free energy as a function of $\bar{k}$ there is a shallow minimum at around $|\bar{k}| \sim 0.7$ that corresponds to ice XI.
This minimum is shifted to higher values of $|\bar{k}|$ as the temperature is lowered as a consequence of reduced thermal fluctuations.
On the other hand, in the free energy as a function of $n_{\mathrm{XI}}$ the ice XI minimum is precisely located at $|n_{\mathrm{XI}}|=96$ at all temperatures.
Furthermore, there is a barrier that separates ice XI from other proton configurations that was not observed when $\bar{k}$ was used as order parameter.
For this reason, we conclude that $n_{\mathrm{XI}}$ provides a greater resolution of the proton configurations in our simulation.
The existence of a barrier separating the two phases is also the hallmark of a first order phase transition, and there is experimental evidence that the transformation is indeed first order\cite{Line96}.
There are also systematic shallow minima that cannot be ascribed to the statistical uncertainty.
These minima are located at $|n_{\mathrm{XI}}| \sim 80$ and $|\bar{k}| \sim 0.6$.
A visual inspection of the configurations shows that they correspond to ordered layers of protons that are likely stabilized by the periodic boundary conditions used in the simulations.

\begin{figure}[t]
\centering
\includegraphics[width=0.6\textwidth]{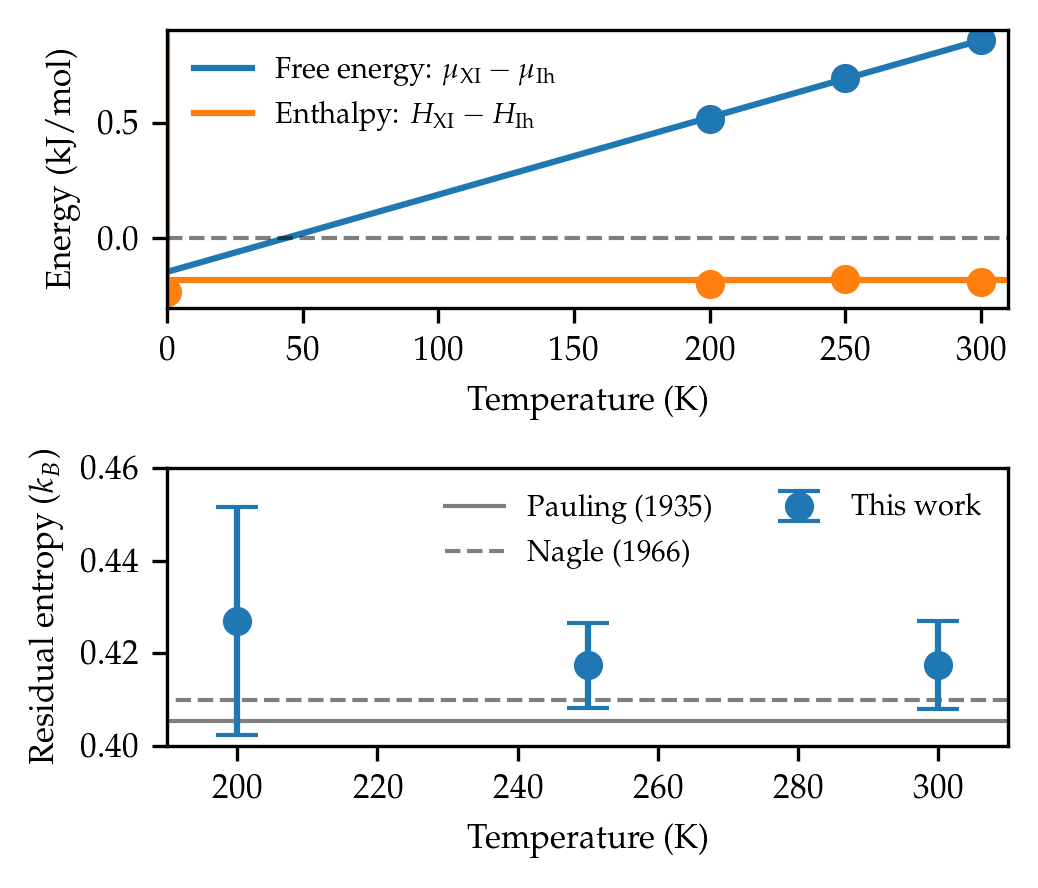}
\caption{\label{fig:plotFE_PE_96} Above) Difference in chemical potential $\mu_{\mathrm{XI}}-\mu_{\mathrm{Ih}}$ and enthalpy $H_{\mathrm{XI}}-H_{\mathrm{Ih}}$ between ice XI and Ih.
The lines were fit to the data at 200, 250, and 300 K.
We have included the enthalpy at 0 K calculated from structure optimizations of ice XI and Ih.
Below) Difference in entropy between ice XI and Ih (residual entropy) calculated from the simulations at difference temperatures. Estimations due to Pauling\cite{Pauling35} and Nagle\cite{Nagle66} are also shown.}
\end{figure}

We also calculated the difference in chemical potential $\mu_{\mathrm{XI}}-\mu_{\mathrm{Ih}}$ and enthalpy $H_{\mathrm{XI}}-H_{\mathrm{Ih}}$ between ice XI and Ih using reweighting.
The results are shown as a function of temperature in Figure \ref{fig:plotFE_PE_96}.
Although our calculations were performed at relatively high temperature, we can extrapolate the results at low temperatures by fitting a straight line to $\mu_{\mathrm{XI}}-\mu_{\mathrm{Ih}}$.
Based on this rough extrapolation, we obtain a transition temperature of around 50 K for the machine learning model based on the SCAN functional employed here.
$H_{\mathrm{XI}}-H_{\mathrm{Ih}}$ seems fairly independent of the temperature and has an average value of 180 J/mol.
This difference in enthalpy is in good agreement with the experimental result 168 J/mol from calorimetric measurements\cite{Tajima84}.
We also calculated $H_{\mathrm{XI}}-H_{\mathrm{Ih}}$ at 0 K by an energy minimization with respect to the atomic coordinates and the box tensor, and obtained $H_{\mathrm{XI}}-H_{\mathrm{Ih}} = 230$  J/mol.
Using this value for the enthalpy and a theoretical estimate for the residual entropy of 0.410 $k_B$ \cite{Nagle66}, a somewhat higher transition temperature of 68 K is obtained.
The agreement with the experimental value of 72 K is particularly good for the latter estimate.
It is instructive to compare our result for the transition temperature with the results obtained with other DFT functionals.
Sch{\"o}nherr \textit{et al.}\cite{Schonherr14} found a transition temperature of 70-80 K, quite close to our estimate, by using a vdW inclusive hybrid version of the Perdew-Burke-Ernzerhof (PBE)\cite{Perdew96} functional.
With the non-hybrid version of their functional they obtained a transition temperature of 90-100 K.
A transition temperature of 105 K was calculated independently by Pamuk \textit{at al.} using the same vdW inclusive PBE functional\cite{Pamuk15}.
Singer \text{et al.}\cite{Singer05} found a transition temperature of 98 K using the Becke-Lee-Yang-Parr\cite{Gill92} (BLYP) functional.
All these results are based on classical calculations.
Pamuk \textit{et al.}\cite{Pamuk15} found that nuclear quantum effects reduce the transition temperature from 105 K to 91 K, for light water, within the quasi-harmonic approximation for the vibrational free energy.  

From the thermodynamic relation $\Delta S=(\Delta H- \Delta G)/T$ we can also calculate the entropy.
In Figure \ref{fig:plotFE_PE_96} we show the residual entropy $S_0$ at the three temperatures studied here.
The residual entropy is independent of temperature within error bars, and the mean value is $S_0=0.417(6)$ $k_B$ if only the simulations at 250 and 300 K are considered.
This value can be compared with Pauling's mean-field result\cite{Pauling35}, $S_0=0.405$ $k_B$, Nagle's  accurate series expansion result\cite{Nagle66}, $S_0=0.410$ $k_B$, and the experimental result\cite{Giauque36}, $S_0=0.41$ $k_B$.
We note that Pauling's result is a lower bound for $S_0$ \cite{Onsager60}, while Nagle's result is an improvement over Pauling's calculation yet neglects enthalpic effects.
Our calculation includes enthalpic effects and agrees with experiment within the error bars, but is affected by the finite size of our cell.
Finite size effects tend to increase the residual entropy, as shown by Herrero and Ramirez\cite{Herrero13} and Berg \textit{et al.}\cite{Berg07}.
Thus, our calculation is likely an upper bound for $S_0$.

We also calculated the polarization $P$ of ice XI along the $c$ crystallographic axis and the dipole moment of  the water molecule in ice XI as a function of temperature.
The results are summarized in Table \ref{tbl:polarization} and are similar to the results of former calculations reported in the literature for ice Ih and ice XI.
\begin{table}[b]
\small
  \caption{Dipole moment $\mu$ of the water molecule and polarization $P$ along the $c$ axis. All results correspond to ice XI and are based on the SCAN functional unless otherwise specified.}
  \label{tbl:polarization}
  \begin{tabular*}{\textwidth}{@{\extracolsep{\fill}}lll}
    \hline
     Temperature (K) & $\mu$ (D) & $P$ ($\mu$C/cm$^2$) \\
    \hline
     0    & 3.41 & 21.64 \\
     50   & 3.39(1) & 21.37(3) \\
     100  & 3.36(1) & 20.99(3) \\
     150  & 3.34(1) & 20.60(5) \\
     200  & 3.31(1) & 20.19(5) \\
     250  & 3.28(1) & 19.75(6) \\
     300  & 3.24(1) & 19.19(7) \\
     0  - PBE \cite{Ishii12} & 3.3 & 21 \\
     273 - Ih \cite{Chen17} & 3.29(21) & - \\
     273 - Ih - PBE \cite{Chen17} & 3.35(21) & - \\
    \hline
  \end{tabular*}
\end{table}

\begin{figure}[t]
\centering
\includegraphics[width=0.6\textwidth]{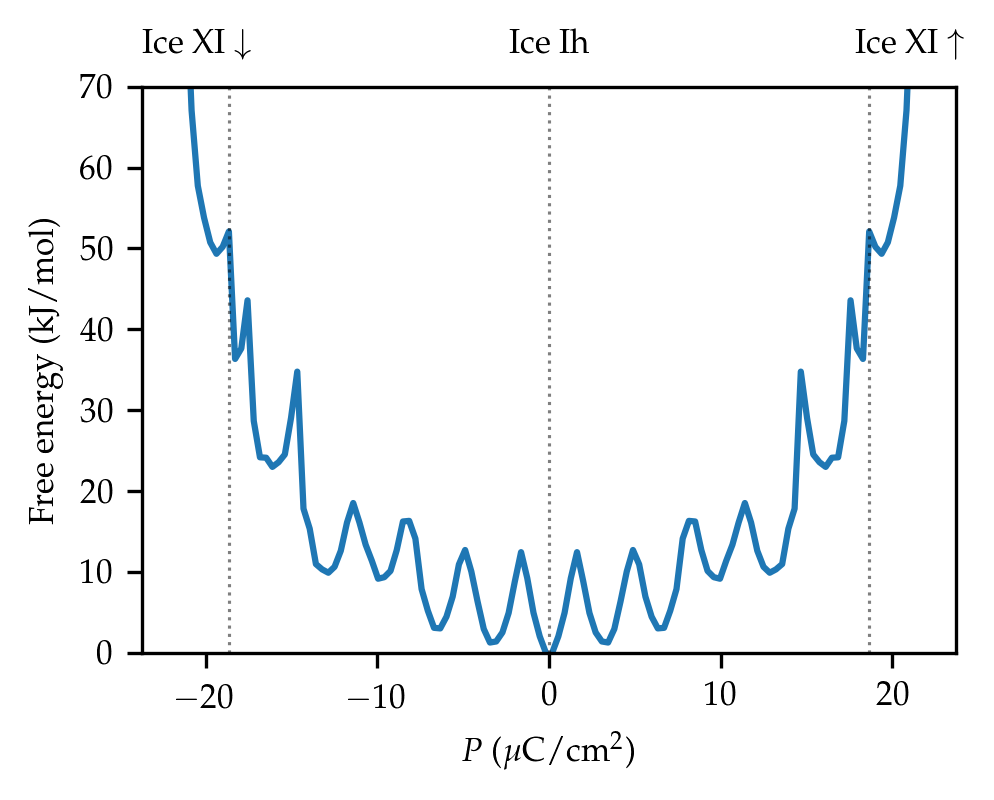}
\caption{\label{fig:polarization} Free energy as a function of polarization $P$ along the $c$ crystallographic axis at 300 K.}
\end{figure}

Since the polarization $P$ is arguably the natural order parameter for the paraelectric-ferroelectric transition, we set out to compute the free energy as a function of $P$ along the $c$ crystallographic axis.
The enhanced sampling simulations were driven by a structural order parameter that depends only on the nuclear coordinates.
However, we can use the reweighting technique in order to compute \textit{a posteriori} how the free energy depends on an order parameter, such as $P$, which includes explicit electronic structure information.  
The free energy as a function of $P$ at 300 K is shown in Figure \ref{fig:polarization}. 
Compared with the plot of the free energy as a function of $n_{\mathrm{XI}}$ or $\bar{k}$ (see Figure \ref{fig:plotLandau}), this plot has more minima corresponding to the different polarization states realized in the course of the transformation. The barriers separating the local minima reflect the activated nature of the rearrangements processes necessary to change the polarization. Comparison of Figure \ref{fig:polarization} with Figure \ref{fig:plotLandau} suggests that $P$ has a higher resolution than the purely structural order parameters used for biasing.  
The lowest free energy minimum is found at $P=0$ and corresponds to the paraelectric phase (ice Ih).
The highest free energy minima are found at around $P=\pm 19.5$ $\mu$C/cm$^2$ and correspond to the ferroelectric phase (ice XI) in its $\uparrow$ and $\downarrow$ variants.
This result is in good agreement with findings using unbiased simulations of ice XI shown in Table \ref{tbl:polarization}.

\section{Conclusions}

We presented a method to accelerate changes in proton configurations in molecular dynamics simulations based on introducing a bias potential that promotes the formation of ionic defects.
This approach provides an alternative to the Monte Carlo algorithm often used to sample proton configurations\cite{Rahman72,Rick03,Barkema98}.
We analyzed the mechanism of the changes in proton configuration.
We found that the initial charge separation in the system is a result of the formation of an \ce{H7O3+} or \ce{H5O2+} (Zundel) cation, and a \ce{OH-} anion.
After the charge separation, most proton jumps along a hydrogen bond loop are a result of the diffusion of the \ce{OH-} anion.
We also found evidence of collective behavior in the diffusion of the \ce{OH-} anion with at least three water molecules involved in a concerted proton jump.
Finally, the two ionic defects recombine and all the protons in the hydrogen bond loop remain flipped.
Changes in the total dipole moment are allowed as a consequence of the periodic boundary conditions that gives rise to winding loops.

We also showed the usefulness of this approach by studying the ice XI/Ih transition.
For this purpose, we constructed an \textit{ad hoc} order parameter for ice XI based on the local atomic environments around oxygen atoms.
An appropriate quasi-static bias potential as a function of this order parameter sufficed to drive reversible transitions between ice XI and ice Ih.
From these simulations we calculated the free energy difference between the two phases and the residual entropy of ice Ih.
We obtained a residual entropy of $0.417(6)$ $k_B$ that is in agreement with the experiment within the error bar of our calculation.
We argue that it represents an upper bound to the true residual entropy since finite size effects reduce the residual entropy.
As far as we know, this is the first calculation of the residual entropy that fully takes into account the subtle variation in energy between different proton configurations.
The approximate transition temperature obtained here from a rough extrapolation of high temperature simulations is around 50 K.
A simpler estimate of the transition temperature based on the enthalpy at 0 K for this model, results in a transition temperature of 68 K in quite good agreement with the experimental result of 72 K, even if we take into account that nuclear quantum effect will likely change our result\cite{Pamuk15}.


Our approach was made possible by the use of a machine learning potential based on the SCAN exchange and correlation functional.
This model is able to describe the dissociation of water and therefore the proton transfer between oxygen atoms.
We found that this model reproduces well the higher stability of ice XI than of Ih at 0 K, and also the ice XI/Ih transition temperature is in reasonable agreement with experiment.
We note that these results depend on the choice of the level of theory used for training the NNP, and that the potential energy surface (PES) described by our model is not a perfect representation of the SCAN PES\cite{Piaggi21}.
It should be borne in mind that our simulations ignore surface and other microstructure effects.
In a slab with free surfaces, charges localized at the surfaces will generate a depolarizing field that may affect the stability of ice XI as discussed in ref.~\citenum{Parkkinen14}.
Here we have ignored such effects and concentrated purely on bulk phenomena, as customary in the modern theory of polarization\cite{Resta94}.
It is an open question which boundary conditions would better reflect the experimental situation.  

The performance of the method presented here to change the proton configuration deteriorates at low temperatures.
Extending the usefulness of this approach to lower temperatures will be the subject of future work.
Furthermore, the order parameter for the ice XI/Ih introduced here can find multiple applications.
For instance, it could be used to mimic the effect of an electric field acting on the system.
This would allow to perform a computer experiment in which the rearrangement of the proton configuration is driven by a simulated external electric field.

\section*{Acknowledgements}
We are grateful to Marcos Calegari for providing the training data for the neural network to predict the positions of the Wannier centroids.
P.M.P was supported by an Early Postdoc.Mobility fellowship from the Swiss National Science Foundation. 
This work was conducted within the center: Chemistry in Solution and at Interfaces funded by the DoE under Award DE-SC0019394.
Simulations reported here were substantially performed using the Princeton Research Computing resources at Princeton University which is consortium of groups including the Princeton Institute for Computational Science and Engineering and the Princeton University Office of Information Technology's Research Computing department.

\end{document}